\documentclass[journal,10pt]{IEEEtran}

\usepackage{color}
\usepackage{multirow}
\usepackage{graphicx}
\usepackage{epstopdf}
\usepackage[cmex10]{amsmath}
\usepackage{amssymb}

\usepackage{multirow}
\usepackage{amsthm}
\usepackage{cite}
\usepackage{acronym} %This package will write out the acronym the first time it is used and write it as an acronym for the rest of the paper
\usepackage{algorithm, algcompatible}

\algnewcommand\INPUT{\item[\textbf{Input:}]}
\algnewcommand\OUTPUT{\item[\textbf{Output:}]}

\allowdisplaybreaks

\begin{document}

\title{Moment-based Spectrum Sensing Under Generalized Noise Channels} 
\author{Nikolaos~I.~Miridakis,~\IEEEmembership{Senior Member,~IEEE}, Theodoros~A.~Tsiftsis,~\IEEEmembership{Senior Member,~IEEE} \\and Guanghua~Yang,~\IEEEmembership{Senior Member,~IEEE}
%\thanks{Copyright \copyright  2019 IEEE. Personal use of this material is permitted. However, permission to use this material for any other purposes must be obtained from the IEEE by sending a request to pubs-permissions@ieee.org.}
%\thanks{\textit{Corresponding Author: T. A. Tsiftsis.}}
\thanks{N. I. Miridakis, T. A. Tsiftsis and G. Yang are with the Institute of Physical Internet and School of Intelligent Systems Science \& Engineering, Jinan University, Zhuhai Campus, Zhuhai 519070, China. N. I. Miridakis is also with the Dept. of Informatics and Computer Engineering, University of West Attica, Aegaleo 12243, Greece (e-mails: nikozm@uniwa.gr, theo\_tsiftsis@jnu.edu.cn, ghyang@jnu.edu.cn).}
}

%\markboth{}{}

\maketitle

\begin{abstract}
A new spectrum sensing detector is proposed and analytically studied, when it operates under generalized noise channels. Particularly, the McLeish distribution is used to model the underlying noise, which is suitable for both non-Gaussian (impulsive) as well as classical Gaussian noise modeling. The introduced detector adopts a moment-based approach, whereas it is not required to know the transmit signal and channel fading statistics (i.e., blind detection). Important performance metrics are presented in closed forms, such as the false-alarm probability, detection probability and decision threshold. Analytical and simulation results are cross-compared validating the accuracy of the proposed approach. Finally, it is demonstrated that the proposed approach outperforms the conventional energy detector in the practical case of noise uncertainty, yet introducing a comparable computational complexity.
\end{abstract}


\IEEEpeerreviewmaketitle

\section{Introduction}
\IEEEPARstart{S}{ignal} detection and spectrum sensing represent two well-known complementary research topics that have attracted a vast research interest over the last decades. Some of the most popular spectrum sensing strategies include the coherent, cyclostationary and energy detection (ED) \cite{j:MariaNI11,j:Mariani19}. The former two approaches are typically optimal under noisy channels; yet with the cost of either requiring perfect synchronization and information of signal statistics or high computational complexity. ED, on the other hand, reflects on a simple implementation as well as it does not require any prior information regarding the transmitted signal and underlying channel fading. However, ED is quite sensitive in noise power uncertainty, which is typically the case in realistic conditions. To this end, another signal detection type has been proposed; namely the moment-based detection (MD), which produces a similar complexity as ED but preserving robustness in the presence of noise uncertainty at the same time \cite{j:Morelande01,c:BogaleVandendorpe13}. However, thus far, MD has been studied under additive white Gaussian noise (AWGN) channels and in the absence of channel fading.    

In practice, there are various types of wireless communication channels where signal transmission is subjected to non-Gaussian (i.e., impulsive) noise. Typical examples include urban and indoor wireless channels, ultra-wide band communications, frequency/time-hopping with jamming, millimeter wave communications, and wireless transmissions under strong interference conditions (see \cite{j:MoghimiNasriSchober11,j:ZhuWang2019} and references therein). Impulsive effects that introduce additive non-Gaussian noise can also be found in cognitive radio (CR); e.g., due to the simultaneous spectrum access under miss-detection events \cite{j:JafarSrinivasa07}. Unfortunately, detectors designed for AWGN do not perform well in non-Gaussian noise \cite{j:YeLi2019,j:QuXu2020}. The most widely-used circularly symmetric non-Gaussian noise models are the Gaussian mixture, Middleton's Class A, $\epsilon$-mixture, Laplacian, generalized Gaussian, and $\alpha$-stable \cite{j:MoghimiNasriSchober11}. The latter model (although denotes an accurate model for impulsive noise) does not have finite moments and thus it is not amenable for further processing. Generally, all the former noise models cannot be suitably interpreted as the sum of a large number of independent and identically distributed impulsive noise sources with small power. Doing so, the lack of matching the real-world phenomena of impulsive noise sources from a Gaussian to a non-Gaussian distribution defines a crucial weakness of the aforementioned non-Gaussian noise distributions. Besides, focusing on the aforementioned non-Gaussian models, existing art studied the generalized energy detection and fractional low-order moment approaches \cite{j:MoghimiNasriSchober11,j:ZhuWang2019}. Nonetheless, these works introduced a considerably high computational complexity and presented critical performance metrics (such as the detection probability or optimal setting parameters) in integral-only forms or by utilizing numerical-search methods. 

On another front, the McLeish distribution {\color{black}(also known as generalized symmetric Laplace or Bessel function distribution)} represents an alternative noise model, appropriate for both Gaussian and non-Gaussian noise channels. It was originated by D. Mcleish in \cite{j:McLeish82} and quite recently it was revisited and thoroughly analyzed in \cite{j:YilmazMcLeish2020,c:YilmazAlouini2018}. McLeish distribution resembles the Gaussian distribution; it is unimodal, symmetric, it has all its moments finite, and has tails that are at least as heavy as those of Gaussian distribution. Moreover, the evolution of its impulsive nature from Gaussian distribution to non-Gaussian distribution is explicitly parameterized in a rigorous way with psychical meaning (please, see the detailed analysis in \cite[\S IV.B.]{j:YilmazMcLeish2020}); especially than those of Laplacian, $\alpha$-stable and generalized Gaussian distributions.

In this letter, for the first time, we study the spectrum sensing and signal detection under McLeish noise channels. A moment-based approach is adopted, aiming to a simple implementation. Also, the considered MD does not have any knowledge of the transmitted signal and channel fading statistics (i.e., blind detection), while it operates on a fast-faded channel. It is only aware of the underlying noise statistics. For sufficiently large number of channel samples, which is usually the practical case, new analytical closed-form expressions are derived regarding some important performance metrics; the false-alarm and detection probabilities as well as the decision threshold. Most importantly, it is explicitly demonstrated that the considered test statistic based on the MD approach is independent of noise power (for both Gaussian and non-Gaussian noise channels), which in turn reflects on a rather robust and accurate detection strategy. In fact, MD is cross-compared to the benchmark ED under noise uncertainty conditions, while the superiority of the former against the latter approach is clearly manifested.  

{\it Notation:} Matrices and vectors are represented by uppercase and lower bold typeface letters, respectively. The coefficient at the $i^{\rm th}$ row and $j^{\rm th}$ column of $\mathbf{A}$ is defined as $\mathbf{A}_{(i,j)}$. Superscript $(\cdot)^{\mathcal{T}}$ denotes vector transposition and $|\cdot|$ represents absolute (scalar) value. $\mathbb{E}[\cdot]$ is the expectation operator and symbol $\overset{\text{d}}=$ means equality in distribution. ${\rm Var}[\cdot]$ and ${\rm Kurt}[\cdot]$ are the variance and kurtosis operators, respectively. $\mu_{x}(n)=\mathbb{E}[x^n]$ represents the $n^{\rm th}$ moment function of a random variable (RV) $x$, while $y_{|z}$ denotes that $y$ is conditioned on $z$ event. $\mathcal{CN}(\mu,\sigma^{2})$ and $\mathcal{N}(\mu,\sigma^{2})$ define, respectively, a complex and circularly symmetric (CCS) Gaussian RV as well as a real-valued Gaussian RV with mean $\mu$ and variance $\sigma^{2}$. Also, $\mathcal{CML}(\mu,\sigma^{2},v)$ and $\mathcal{ML}(\mu,\sigma^{2},v)$ denote, respectively, a CCS and real-valued RV following the McLeish distribution with mean $\mu$, variance $\sigma^{2}$ and non-Gaussianity parameter $v$. Further, $Q(\cdot)$ and $Q^{-1}(\cdot)$ are the Gaussian $Q$-function and inverse $Q$-function, respectively. $\Gamma(\cdot)$ denotes the Gamma function \cite[Eq. (8.310.1)]{tables}, $(\cdot)!!$ is the double factorial operator \cite[p. xliii]{tables} and $K_{v}(\cdot)$ denotes the $v^{\rm th}$ order modified Bessel function of the second kind \cite[Eq. (8.432)]{tables}. Finally, ${\rm Re}\{x\}$ and ${\rm Im}\{x\}$ denote the real and imaginary part of a complex-valued $x$, respectively.

\section{System and Signal Model}
Consider the classical binary hypothesis problem applied on a wireless spectrum sensing device, which reads as
\begin{align}
\begin{array}{l l l}     
    \mathcal{H}_{0}: &y[u]=w[u],& \text{no signal is present,} \\
    \mathcal{H}_{1}: &y[u]=h[u] s[u]+w[u],  & \text{signal transmission,}
\end{array}
\label{proform}
\end{align}
where $y[u] \in \mathbb{C}$, $h[u]\in \mathbb{C}$, $s[u] \in \mathbb{R}$ and $w[u] \in \mathbb{C}$ denote the received signal, channel fading coefficient, transmitted discrete-time baseband signal and additive noise, respectively, at the $u^{\rm th}$ sample. The transmitted signal samples, $s[\cdot]$, are captured by a given constellation with transmit power $s_{\rm p}$, whereby can be efficiently modeled as discrete uniformly distributed RVs. In most practical wireless digital applications, such a constellation may be either an $\mathcal{M}$-ary phase shift keying ($\mathcal{M}$-PSK) or quadrature amplitude modulation ($\mathcal{M}$-QAM). Further, it is assumed that the channel fading coefficient, $h[\cdot]$, may follow an arbitrary distribution. {\color{black}Also, $h[i]\neq h[j]$ and $w[i]\neq w[j]$ $\forall i\neq j$, while $h[\cdot]$ and $w[\cdot]$ remain unchanged during a sample duration, whereas they may change between consecutive samples.} 

Additionally, $w[\cdot]\overset{\text{d}}=\mathcal{CML}(0,\sigma^{2}_{w},v)$ with $\sigma^{2}_{w}\in \mathbb{R}^{+}$ and $v\in \mathbb{R}^{+}$ standing for the noise variance and non-Gaussianity parameter, respectively, with a symmetric and unimodal probability density function defined as \cite[Eq. (85)]{j:YilmazMcLeish2020}
\begin{align}
f_{w}(w)=\frac{2 \sqrt{v} |w|^{v-1}}{\sqrt{2 \sigma^{2}_{w}} \pi \Gamma(v)}K_{v-1}\left(\sqrt{\frac{2 v}{\sigma^{2}_{w}}} |w|\right).
\label{noisePDF}
\end{align}
Some special cases of $f_{w}(\cdot)$ are obtained for $v=1$, $v\rightarrow +\infty$ and $v\rightarrow 0^{+}$ resulting to the CCS Laplacian, Gaussian and Dirac's distribution, respectively \cite{j:YilmazMcLeish2020}. It turns out that the McLeish distribution is a generalized and versatile distribution model, which is suitable for both Gaussian and non-Gaussian (impulsive) noise channels. In fact, the noise statistics can be computed and fitted to the McLeish distribution model\footnote{Illustratively, IEEE 802.22 and ECMA 392 standards utilize sporadic long sensing periods for fine sensing and more frequent short sensing periods in which a variety of signal-free samples can be collected and further processed for noise estimation \cite{j:SenanayakeSmith20}.} using
\begin{align}
\sigma^{2}_{w}\triangleq {\rm Var}[w],\quad \text{and} \quad v\triangleq \frac{3}{{\rm Kurt}[w]-3}.
\label{stats}
\end{align}

Moreover, it is assumed that the spectrum sensing device is fully unaware of both the instantaneous and statistical channel gains as well as the signal statistics (i.e., neither the variance of channel gains nor the transmit signal power and the utilized modulation scheme are available); reflecting on a \emph{blind} spectrum sensing. Yet, it is assumed that the noise statistics are known. Thereby, a moment-based estimator is adopted for the test statistic including the 4$^{\text{th}}$ and 2$^{\text{nd}}$ absolute moments of the received signal, which reads as \cite{j:Morelande01}
\begin{align}
T\triangleq -\frac{\mu_{|y|}(4)}{\mu^{2}_{|y|}(2)}.
\label{momest}
\end{align}
Note that ${\rm Kurt}[|y|]=-T$. For the $\mathcal{H}_{0}$ hypothesis, we simply get
\begin{align}
T_{|\mathcal{H}_{0}}= -\frac{\mathbb{E}[|w|^{4}]}{\mathbb{E}[|w|^{2}]^{2}}=-\left(2+\frac{3}{2 v}\right).
\label{TestH0}
\end{align}
The proof of \eqref{TestH0} is provided in Appendix \ref{appa} with the aid of \eqref{mom2} and \eqref{mom4} and after some simple manipulations. It is noted that $T \rightarrow -2$ as $v \rightarrow +\infty$ (i.e., for AWGN channels), which is in accordance to \cite{j:Morelande01}, while $T=-7/2$ for Laplacian noise. It turns out that the sensing problem can be formulated by setting that the considered test is equal or greater than $-2-3/(2 v)$, reflecting the signal absence or presence, correspondingly, yielding
\begin{align}
T\overset{\mathcal{H}_{1}}{\underset{\mathcal{H}_{0}}{\geqq}}-\left(2+\frac{3}{2 v}\right).
\label{TestGeneralState}
\end{align}

\section{Performance Metrics}
In realistic conditions, $T$ is obtained by estimating the moments function of the received signal via a given number of samples, $N$, such that 
\begin{align}
\hat{T}= -\frac{\hat{\mu}_{|y|}(4)}{\hat{\mu}^{2}_{|y|}(2)},
\label{SampleTest}
\end{align}
where
\begin{align}
\hat{\mu}_{|y|}(n)\triangleq \frac{1}{N}\sum^{N}_{u=1}|y[u]|^{n}.
\label{SampleMom}
\end{align}
Obviously, $\hat{T}\rightarrow T$ as $N\rightarrow +\infty$. To obtain the exact test statistics and to evaluate the mismatch between actual and estimated moments, we introduce the RV $\sqrt{N}(\hat{T}-T)$, which has the following asymptotic property based on the central limit theorem (CLT):
\begin{align}
\sqrt{N}(\hat{T}-T)\overset{\text{d}}= \mathcal{N}\left(0,\sigma^{2}\right),
\label{RVTest}
\end{align}
where 
\begin{align}
\sigma^{2}_{|\mathcal{H}_{0}}=\frac{16v^{3}+120 v^{2}+294 v+189}{4 v^{3}},
\label{sigmaH0}
\end{align}
and
\begin{align}
\nonumber
\sigma^{2}_{|\mathcal{H}_{1}}=&\bigg[2 \mu^{6}_{x}(2)-4 \mu_{x}(4) \mu^{4}_{x}(2)+\left(\mu^{2}_{x}(4)+\mu_{x}(8)\right)\mu^{2}_{x}(2)\\
&-4 \mu_{x}(4) \mu_{x}(6) \mu_{x}(2)+4 \mu^{3}_{x}(4)\bigg] \left(8 \mu^{6}_{x}(2)\right)^{-1},
\label{sigmaH1}
\end{align}
with symbol $x$ standing for a shorthand notation for ${\rm Re}\{y\}$ and $\mu_{x}(n)=\mu_{{\rm Re}\{y\}}(n)$ is given in \eqref{momML}. The proof is relegated in Appendix \ref{appb}. It is noteworthy that $\sigma^{2}_{|\mathcal{H}_{0}}=4$ in the AWGN case (when $v\rightarrow +\infty$), as it should be \cite{j:Morelande01}. Also, an insightful remark is the fact that \eqref{TestH0} and \eqref{sigmaH0} are independent of noise power $\sigma^{2}_{w}$ under hypothesis $\mathcal{H}_{0}$; reflecting on quite an efficient and robust test statistic in the presence of detrimental yet realistic uncertain noise power estimation \cite{j:Tandra}. 

The scenario of a false-alarm probability, namely, $P_{f}(\cdot)$, is formulated as $P_{f}(\lambda)\triangleq \text{Pr}[\hat{T}>\lambda|\mathcal{H}_{0}]$ with $\lambda$ denoting the decision threshold. Hence, since $\hat{T}_{|\mathcal{H}_{0}}\overset{\text{d}}= \mathcal{N}(0,\sigma^{2}_{|\mathcal{H}_{0}})$, it holds that
\begin{align}
\nonumber
P_{f}(\lambda)&=Q\left(\lambda/\sigma_{|\mathcal{H}_{0}}\right)\\
&=Q\left(\frac{\lambda \sqrt{4 v^{3}}}{\sqrt{16v^{3}+120 v^{2}+294 v+189}}\right).
\label{Pf}
\end{align}

As it is obvious from (\ref{Pf}), the false-alarm probability is an \emph{offline} operation, i.e., it is independent of the instantaneous channel gain, the presence of signal transmission as well as the noise power $\sigma^{2}_{w}$. For known $N$, the common practice of setting the decision threshold is based on the constant false-alarm probability. Also, this is a reasonable assumption since for various practical spectrum sensing applications, the highest priority is to satisfy a predetermined false-alarm rate (e.g., underlay CR). Doing so, the desired threshold, $\lambda^{\star}$, yields as
\begin{align}
\lambda^{\star}\triangleq Q^{-1}\left(P^{({\tau})}_{f}\right) \frac{(16v^{3}+120 v^{2}+294 v+189)}{4 v^{3}},
\label{thr}
\end{align} 
where $P^{({\tau})}_{f}$ represents the predetermined target on the maximum attainable false-alarm probability.

In the case of signal transmission, modeled by the $\mathcal{H}_{1}$ hypothesis, the estimated test statistic is distributed as
\begin{align}
\hat{T}\overset{\text{d}}=\mathcal{N}\left(\sqrt{N}\left(T+2+\frac{3}{2 v}\right),\sigma^{2}_{|\mathcal{H}_{1}}\right).
\label{testH1distr}
\end{align} 
Thus, the detection probability, $P_{d}(\cdot)$, is directly obtained in a closed form as
\begin{align}
\nonumber
P_{d}(\lambda)&\triangleq \text{Pr}[\hat{T}>\lambda|\mathcal{H}_{1}]\\
&=Q\left(\frac{\lambda^{\star}-\sqrt{N}\left(T_{|\mathcal{H}_{1}}+2+\frac{3}{2 v}\right)}{\sigma_{|\mathcal{H}_{1}}}\right),
\label{Pd}
\end{align} 
where $T_{|\mathcal{H}_{1}}$ is the estimation test during the $\mathcal{H}_{1}$ hypothesis expressed as
\begin{align}
%\nonumber
T_{|\mathcal{H}_{1}}=-\frac{\mu_{|y|}(4)}{\mu^{2}_{|y|}(2)}= -\frac{\mu_{{\rm Re}\{y\}}(4)}{2 \mu^{2}_{{\rm Re}\{y\}}(2)}-\frac{1}{2},
\label{TH1}
\end{align} 
with $\mu_{{\rm Re}\{y\}}(n)$ provided by \eqref{momML}. 

\section{Numerical Results and Discussion}
In this section, the derived analytical results are verified via numerical validation where they are cross-compared with corresponding Monte-Carlo simulations. For the signal transmission, it is assumed that both the transmitter and receiver employ a square-root raised cosine
filter with roll-off factor $0.2$, oversampling factor $F=4$ and filter length $4 F+1$. Also, zero-mean CCS McLeish noise instances, i.e., $\mathcal{CML}(0,\sigma^{2}_{w})$, are generated as per \cite[Thm. 10]{j:YilmazMcLeish2020}, while it is assumed that the non-Gaussianity parameter $v$ is known at the receiver. In what follows, and without loss of generality, $h[u]\overset{\text{d}}=\mathcal{CN}(0,1)$ for the $u^{\rm th}$ sample; reflecting on unit-scale Rayleigh fast-faded channels. Hence, the received signal-to-noise ratio (SNR) is defined as $\text{SNR}\triangleq s^{2}_{\rm p}/\sigma^{2}_{w}$. All the simulation results are conducted by averaging $10^{4}$ independent trials. Hereinafter, line-curves and cross-marks denote the analytical and simulation results, respectively.   

\begin{figure}[!t]
\centering
\includegraphics[trim=1.5cm 0.2cm 0.5cm .1cm, clip=true,totalheight=0.25\textheight]{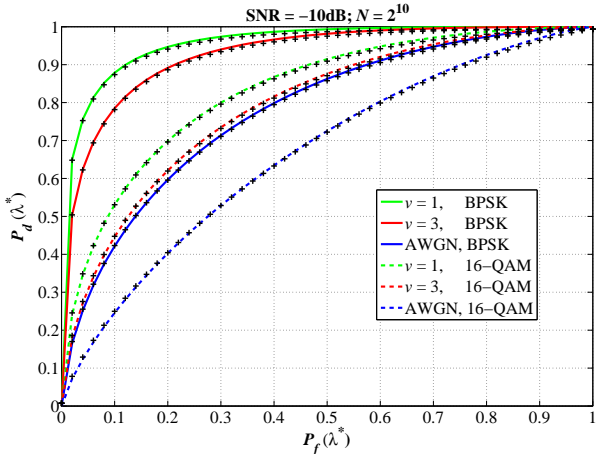}
\caption{Detection probability vs. false-alarm probability for different modulation schemes and noise channels.}
\label{fig1}
\end{figure}

In Fig.~\ref{fig1}, the receiver operating characteristic (ROC) curve is illustrated for the case when BPSK (i.e., $\mathcal{M}=2$) or 16-QAM (i.e., $\mathcal{M}=4$) modulation scheme is applied for data transmission. Also, the CCS Laplacian ($v=1$) and Gaussian ($v\rightarrow +\infty$) noise models are included as the two extreme scenarios. Interestingly, the detection performance is being enhanced for lower-order modulation and/or when the noise becomes non-Gaussian and thus more impulsive (reflecting on a reduced value of $v$). This is an insightful outcome because the increased robustness on non-Gaussian noise is similar to the performance of the detectors based on fractional low-order moment statistics \cite{j:ZhuWang2019}; yet utilizing considerably lower computational efforts. 

In Fig.~2, MD is compared to ED in low SNR regions. For a fair comparison, the decision threshold of the energy detector as well as its corresponding performance metrics should be carefully designed in McLeish noise channels; i.e., see Appendix~\ref{appc}. Additionally, the practical scenario of uncertain estimation of noise power is considered by assuming a uniformly distributed (in dB) uncertainty factor $\beta\triangleq \frac{\hat{\sigma}^{2}_{w}}{\sigma^{2}_{w}}$, where $\hat{\sigma}^{2}_{w}$ is the estimated noise power, while $\beta \in [-L,L]$ with $L\leq 2$dB \cite{j:Tandra}. Clearly, the performance difference between the moment-based detector and the conventional energy detector is emphatic in the presence of noise uncertainty. Although the ED performance gets worse for impulsive non-Gaussian noise (as expected; and in accordance to \cite[Fig.~6]{j:ZhuWang2019}), the MD performance is being enhanced which verifies the aforementioned outcomes from Fig.~\ref{fig1}.  

\begin{figure}[!t]
\centering
\includegraphics[trim=1.5cm 0.2cm 0.5cm .1cm, clip=true,totalheight=0.25\textheight]{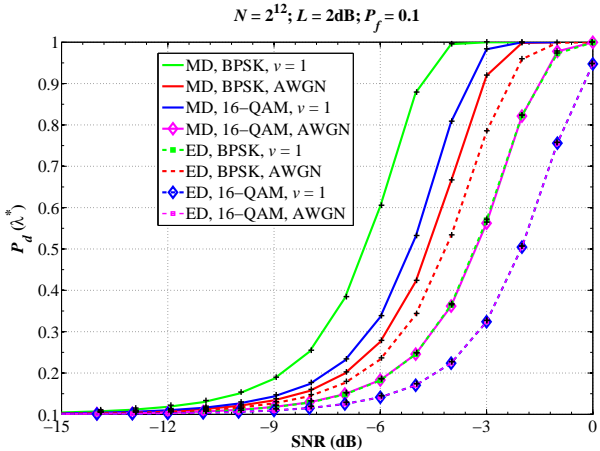}
\caption{Detection probability vs. various SNR values for different modulation schemes and noise channels. MD is cross-compared to ED.}
\label{fig2}
\end{figure}

\begin{figure}[!t]
\centering
\includegraphics[trim=1.5cm 0.2cm 0.5cm .1cm, clip=true,totalheight=0.25\textheight]{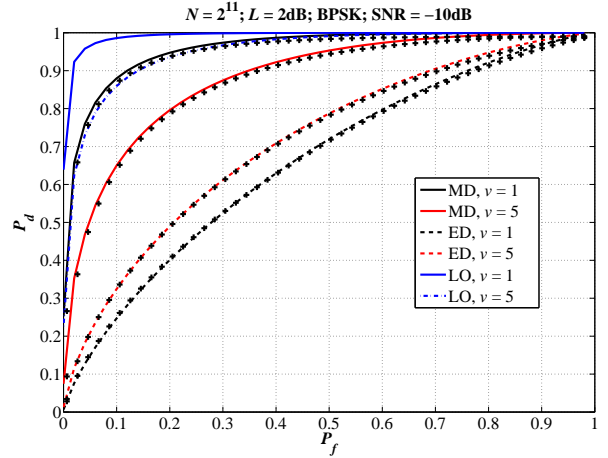}
\caption{Detection probability vs. false-alarm probability for different detectors and noise channels.}
\label{fig3}
\end{figure}

{\color{black}In a similar basis, the ROC performance of three different detectors is illustrated in Fig.~\ref{fig3}. In particular, the considered MD is compared with ED as well as the locally optimal (LO) detector \cite[Eq. (9)]{j:MoghimiNasriSchober11} which is numerically computed. Note that the LO detector approaches the global optimal in low SNR regions and, hence, may serve as a performance benchmark. In accordance to the aforementioned discussion, MD outperforms ED in the presence of noise uncertainty, while the said performance difference increases for more impulsive noise channels (with a reduced $v$). It is also obvious that LO outperforms both MD and ED under non-Gaussian noise; yet, at the cost of considerably higher computational efforts as compared to MD and ED.}

\section{Concluding Remarks}
Blind spectrum sensing and signal detection was studied under CCS McLeish noise, fast-faded channels and discrete $\mathcal{M}$-ary PSK or $\mathcal{M}$-ary QAM signals. A moment-based approach was considered to implement the test statistic, which generalizes the AWGN case to the impulsive non-Gaussian noise environments. Insightfully, the discussed test statistic based on MD is independent of noise power (and hence noise uncertainty), which corresponds to a relatively simple yet quite accurate and robust detection strategy; especially in channels with potentially many unpredicted impulsive sources. Finally, the superiority of MD against ED was demonstrated, while the performance of the former approach is being enhanced for more impulsive noise channels.  

\appendix
\subsection{Derivation of Moments Functions}
\label{appa}
\numberwithin{equation}{subsection}
\setcounter{equation}{0}
The $n^{\text{th}}$ moment function of a discrete uniform RV with $\mathcal{M}$ possible values in the range $[-s_{\rm p},s_{\rm p}]$ is given by
\begin{align}
\mu_{s}(n)=\sum^{\mathcal{M}-1}_{l=0}\frac{(-1)^{n} (\mathcal{M}-2 l-1)^{n} s^{n}_{\rm p}}{\mathcal{M}(\mathcal{M}-1)^{n}},\quad n\in \mathbb{R}^{+}.
\label{momU}
\end{align}
For a CCS $h$, ${\rm Re}\{h\}\overset{\text{d}}={\rm Im}\{h\}\overset{\text{d}}=\mathcal{N}(0,\sigma^{2}_{h}/2)$. Then, the $n^{\text{th}}$ moment function of ${\rm Re}\{h\}$ reads as
\begin{align}
\mu_{{\rm Re}\{h\}}(n)=(n-1)!! \frac{(1+(-1)^{n})\sigma^{n}_{h}}{4},\quad n\in \mathbb{R}^{+}.
\label{momH}
\end{align}
Further, ${\rm Re}\{y\}={\rm Re}\{h s\}+{\rm Re}\{w\}\overset{\text{d}}={\rm Im}\{y\}$. The $n^{\text{th}}$ moment function of ${\rm Re}\{y\}$ is derived by \cite[Eq. (30)]{j:YilmazMcLeish2020}
\begin{align}
\nonumber
\mu_{{\rm Re}\{y\}}(n)&=\sum^{n}_{k=0}\binom{n}{k}\frac{\left(1+(-1\right)^{k})\Gamma\left(v+\frac{k}{2}\right)\Gamma\left(\frac{1+k}{2}\right)}{2 \Gamma\left(v\right) \Gamma\left(\frac{1}{2}\right)}\\
&\times \left(\frac{\sigma^{2}_{w}}{v}\right)^{\frac{k}{2}} \mu_{{\rm Re}\{h\}}(n-k) \mu_{s}(n-k),\quad n\in \mathbb{N}^{+}.
\label{momML}
\end{align}
For the $\mathcal{H}_{1}$ hypothesis (i.e., signal transmission), while utilizing \eqref{momU}-\eqref{momML}, the $2^{\text{nd}}$ absolute moment of the received signal, $y$, is computed in a closed form by 
\begin{align}
\nonumber
\mu_{|y|}(2)&=\mathbb{E}\left[|h s+w|^{2}\right]\\
\nonumber
&=\mathbb{E}\left[\left({\rm Re}\{h s\}+{\rm Re}\{w\}\right)^{2}+\left({\rm Im}\{h s\}+{\rm Im}\{w\}\right)^{2}\right]\\
&=2 \mu_{{\rm Re}\{y\}}(2).
\label{mom2}
\end{align}
Likewise, we have that
\begin{align}
\nonumber
&\mu_{|y|}(4)=\mathbb{E}\left[|h s+w|^{4}\right]\\
\nonumber
&=\mathbb{E}\left[\left(\left({\rm Re}\{h s\}+{\rm Re}\{w\}\right)^{2}+\left({\rm Im}\{h s\}+{\rm Im}\{w\}\right)^{2}\right)^{2}\right]\\
&=2 \left(\mu_{{\rm Re}\{y\}}(4)+\mu^{2}_{{\rm Re}\{y\}}(2)\right).
\label{mom4}
\end{align}
In a similar basis, we get
\begin{align}
\mu_{|y|}(6)=2 \mu_{{\rm Re}\{y\}}(6)+6 \mu_{{\rm Re}\{y\}}(4) \mu_{{\rm Re}\{y\}}(2),
\label{mom6}
\end{align}
and
\begin{align}
\mu_{|y|}(8)=2 \mu_{{\rm Re}\{y\}}(8)+8 \mu_{{\rm Re}\{y\}}(6) \mu_{{\rm Re}\{y\}}(2)+6 \mu^{2}_{{\rm Re}\{y\}}(4).
\label{mom8}
\end{align}
Regarding the $\mathcal{H}_{0}$ hypothesis (i.e., absence of signal $s$), the latter absolute moments of $y$ are directly computed by setting $k=n$ in \eqref{momML}.

\subsection{Derivation of $\sigma^{2}$}
\label{appb}
\numberwithin{equation}{subsection}
\setcounter{equation}{0}
According to \cite[Thm. 3.3.A]{b:Serfling80}, the RV defined in \eqref{RVTest} is a zero-mean Gaussian RV with variance $\sigma^{2}\triangleq \mathbf{c}\mathbf{\Sigma}\mathbf{c}^{\mathcal{T}}$, such that
\begin{align}
\nonumber
\mathbf{c}&=\left[\frac{\partial \hat{T}}{\partial \hat{\mu}_{|y|}(2)},\frac{\partial \hat{T}}{\partial \hat{\mu}_{|y|}(4)}\right]_{\hat{\mu}_{|y|}(2)=\mu_{|y|}(2),\hat{\mu}_{|y|}(4)=\mu_{|y|}(4)}\\
&=\left[\frac{2 \mu_{|y|}(4)}{\mu^{3}_{|y|}(2)},-\frac{1}{\mu^{2}_{|y|}(2)}\right],
\label{cvec}
\end{align}
and
\begin{align}
\mathbf{\Sigma}=\left[{\begin{array}{l l}
   \mu_{|y|}(4)-\mu^{2}_{|y|}(2) & \mu_{|y|}(6)-\mu_{|y|}(2) \mu_{|y|}(4) \\
   \mu_{|y|}(6)-\mu_{|y|}(2) \mu_{|y|}(4) & \mu_{|y|}(8)-\mu^{2}_{|y|}(4) \\
  \end{array} }\right].
\label{Smatrix}
\end{align}
It follows that
\begin{align}
\nonumber
&\sigma^{2}=\\
&\frac{4 \mu^{2}_{|y|}(4) \mathbf{\Sigma}_{(1,1)}-4 \mu_{|y|}(4) \mu_{|y|}(2) \mathbf{\Sigma}_{(1,2)}+\mu^{2}_{|y|}(2) \mathbf{\Sigma}_{(2,2)}}{\mu^{6}_{|y|}(2)}.
\label{sigmaAnalytic}
\end{align}
Consequently, utilizing the relevant moments formulae from Appendix \ref{appa} and after performing some tedious yet straightforward manipulations, we arrive at \eqref{sigmaH0} and \eqref{sigmaH1}, correspondingly.

\subsection{Statistics for Energy Detection}
\label{appc}
\numberwithin{equation}{subsection}
\setcounter{equation}{0}
The normalized test statistic of the typical energy detector is defined as \cite{j:MoghimiNasriSchober11,j:SenanayakeSmith20} 
\begin{align}
T_{\rm ED}\triangleq \frac{1}{N \sigma^{2}_{w}}\sum^{N}_{u=1}|y[u]|^{2}. 
\label{ED}
\end{align}
For sufficiently large number of mutually independent samples, $N$, while invoking CLT, the latter statistic is approximately distributed as $\mathcal{N}(\mathbb{E}[|y|^{2}]/\sigma^{2}_{w},{\rm Var}[|y|^{2}]/(N \sigma^{4}_{w}))$. Then, it is straightforward to show that
\begin{align}
P_{f}(\lambda_{\rm ED})=Q\left(\frac{\lambda_{\rm ED}-\mathbb{E}[|y|^{2}_{|\mathcal{H}_{0}}]/\sigma^{2}_{w}}{\sqrt{{\rm Var}[|y|^{2}_{|\mathcal{H}_{0}}]/(N \sigma^{4}_{w})}}\right),
\label{PfEenrgy}
\end{align}
and
\begin{align}
P_{d}(\lambda_{\rm ED})=Q\left(\frac{\lambda_{\rm ED}-\mathbb{E}[|y|^{2}_{|\mathcal{H}_{1}}]/\sigma^{2}_{w}}{\sqrt{{\rm Var}[|y|^{2}_{|\mathcal{H}_{1}}]/(N \sigma^{4}_{w})}}\right),
\label{PdEenrgy}
\end{align}
where $\lambda_{\rm ED}$ is the decision threshold under energy detection, which is specified in a similar basis as per \eqref{thr} by $\lambda^{\star}_{\rm ED}\triangleq \sqrt{{\rm Var}[|y|^{2}_{|\mathcal{H}_{0}}]/(N \sigma^{4}_{w})} Q^{-1}(P^{({\tau})}_{f})+\mathbb{E}[|y|^{2}_{|\mathcal{H}_{0}}]/\sigma^{2}_{w}$.
Note that $\mathbb{E}[|y|^{2}]=\mu_{|y|}(2)$ and ${\rm Var}[|y|^{2}]=\mu_{|y|}(4)-\mu^{2}_{|y|}(2)$ for either $\mathcal{H}_{0}$ or $\mathcal{H}_{1}$. Without delving into details, the above expressions can be easily computed in closed forms via the relevant formulae of Appendix \ref{appa}, namely, \eqref{mom2}, \eqref{mom4} in conjunction with \eqref{momML}.

\bibliographystyle{IEEEtran}
\bibliography{IEEEabrv,References}

\vfill

\end{document}